\documentclass[]{iopart}
\usepackage{iopams}
\usepackage{setstack}
\usepackage[]{graphicx}
\usepackage{color}
\usepackage[latin1]{inputenc}

\newcommand{\bj}{\bi{j}}
\newcommand{\bv}{\bi{v}}
\newcommand{\bE}{\bi{E}}
\newcommand{\bB}{\bi{B}}
\newcommand{\cK}{{\cal{K}}}

\newcommand{\cE}{{\cal{E}}}
\newcommand{\cU}{{\cal{U}}}
\newcommand{\bcB}{\boldsymbol{\cal{B}}}
\newcommand{\bcE}{\boldsymbol{\cal{E}}}

\begin{document}

\title[Watt-balance: force and voltage on a conductor in a magnetic field]{The watt-balance operation:\\ magnetic force and induced electric potential on a conductor in a magnetic field}
\author{C P Sasso E Massa and G Mana}
\address{INRIM -- Istituto Nazionale di Ricerca Metrologica, str.\ delle Cacce 91, 10135 Torino, Italy}

\begin{abstract}
In a watt balance experiment, separate measurements of magnetic force and induced electric potential in a conductor in a magnetic field allow for a virtual comparison between mechanical and electrical powers, which leads to and an accurate measurement of the Planck constant. In this paper, the macroscopic equations for the magnetic force and the induced electric potential are re-examined from a microscopic point of view and the corrective terms due to a non-uniform density of the conduction electrons induced by their interaction with the magnetic field are investigated. The results indicate that these corrections are irrelevant to the watt balance operation.
\end{abstract}

\submitto{Metrologia}

\pacs{06.20.Jr, 03.50.De, 47.65.-d}

\eads{c.sasso@inrim.it}

\section{Introduction}
The force $F=BLI$ acting on a wire of length $L$, constrained to be at rest in a magnetic flux density $B$ and carrying the electrical current $I$ orthogonal to $B$, is derived by integrating the Lorentz force on the conduction electrons, under the assumptions of current and field uniformity \cite{Fletcher:2003}. Similarly, the electric potential $\cU=BLu$ induced on a wire of length $L$, moving at the velocity $u$ orthogonal to $B$, is derived by assuming an uniform charge-carrier density. If the force $F$ counterbalances the weight $mg$ of a mass $m$ in the gravitational field $g$, by combining these equations and eliminating the geometric factor $BL$, we obtain the equation $mgu=\cU I$, which virtually relates mechanical and electrical powers and allows $m$ to be determined in terms of electrical quantities and, hence, of the Planck constant.

A number of subtleties have been dismissed in the previous analysis. Firstly, the Lorentz force acts on the free electrons, but the forces of constraint act on the ion lattice. The microscopic origin of the magnetic force has been investigated by many authors in order to resolve apparent inconsistencies \cite{Rostoker:1952,Mosca:1974,Redinz:2011}. The conclusion is that the Hall field is the means whereby the force on the conduction electrons is transferred to the positively charged ions. This conclusion avoids misinterpretations, where a magnetic field does a work on the conductor. It has also been shown that the Hall field leads to the magnetic force $BLI$ without violating the identity between the electrical and mechanical powers, once the Joule power dissipated in the conductor has been taken into account. Corrections to the Ampere force-law were also proposed \cite{Goedecke:2006,Sakai:2010}.

In the second place, the Hall field -- together with the electron- and ion-plasma stiffness -- makes the electron and ion densities non-uniform and it induces charge layers close to the wire surfaces orthogonal to the Hall field. Consequently, the electrical-current density is not uniform.

In the third place, in a moving conductor, the Lorentz force strains both the electron gas and the ions thus originating a compensating electric field. But, since the Lorentz force is counteracted also by the electrons and ions elasticity, the induced electric potential is not as high as expected.

These phenomena impose a reanalysis of the  $mgu=\cU I$ equation. Therefore, we derived the magnetic force and the induced electric potential by using a magneto-hydrodynamical model proposed in \cite{Goedecke:2006}. Accordingly, the watt-balance coil is described by two overlapping compressible charged fluids, which are coupled to the external and self-induced electric and magnetic fields. In both the cases, a coil either carrying an electrical current or moving in the magnetic field, this model predicts non-uniform charge distributions that are the sources of electrical fields in a direction transverse to both the magnetic field and the charge motions. Our study was prompted by the discrepancy between the Planck constant values measured in different watt-balance experiments \cite{Steiner:2005,Steele:2012}; it is also a preliminary step to understand the physics of a proposed cryogenic version of the experiment \cite{Picard:2007}.

\section{Watt balance operation}
The watt-balance experiment virtually compares the mechanical and the electrical powers produced by the motion of a mass $m$ in the earth gravitational field and by the motion of the supporting coil in a magnetic field, respectively \cite{Steiner:2005,Robinson:2012}. The comparison is carried out in two steps.

\begin{figure}
\centering
\includegraphics[width=70mm]{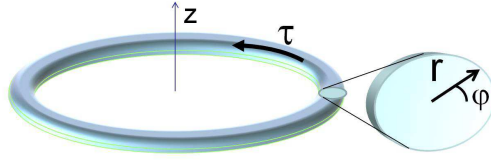}
\centering
\caption{The pseudo-cylindrical coordinate system used in (\ref{static:1}) and (\ref{dynamic:1}); $\tau$ runs over the wire length, $r$ and $\varphi$ are polar coordinates in the wire cross-section.}\label{fig:coord_system}
\end{figure}

Firstly, a balance is used to compare the weight $m g$ with the force generated by the interaction between the electrical current $I$ flowing in the coil and the magnetic flux density $\boldsymbol{B}$. Hence, by using the pseudo-cylindrical coordinate system defined in Fig.\ \ref{fig:coord_system}, this balance is expressed as
\begin{equation}\label{static:1}
m g - \boldsymbol{\hat z} \cdot \int_0^L \int_0^{2\pi} \int_0^{r_0}
 r \boldsymbol{j}(r,\varphi,\tau) \times \boldsymbol{B}(r,\varphi,\tau)\, \rmd r\, \rmd \varphi\, \rmd \tau = 0,
\end{equation}
where $\boldsymbol{\hat z}$ is the vertical direction, $-g\hat{\bi{z}}$ is the gravitational field, $\boldsymbol{j}(r,\varphi,\tau)$ is the current density, $r$, $\varphi$, and $\tau$ are pseudo-cylindrical coordinates along the coil wires, ${r_0}$ is the wire radius and $L$ is its length.

In the second step, the coil is moved and the electric potential induced at the coil ends,
\begin{equation}\label{dynamic:1}\fl
 \cU = \int_0^L \big[ \boldsymbol{u}(r,\varphi,\tau) \times \boldsymbol{B}(r,\varphi,\tau) \big]
 \cdot \hat{\boldsymbol{\tau}} \, \rmd \tau
 =  \int_0^L \boldsymbol{u}(r,\varphi,\tau) \cdot \big[ \boldsymbol{B}(r,\varphi,\tau) \times \hat{\boldsymbol{\tau}} \big]
 \, \rmd \tau ,
\end{equation}
is measured. It must be noted that, provided the end surfaces $\tau=0$ and $\tau=L$ are equipotential, $\cU$ is independent of the integration path. Therefore, we can set $r=0$ and evaluate (\ref{dynamic:1}) on the $r=0$ coil axis.

If $\boldsymbol{j}=I\hat{\boldsymbol{\tau}}/(2\pi r_0^2)$ and $\boldsymbol{B}$ are uniform and the coil velocity is the same everywhere and parallel to $\boldsymbol{\hat z}$, that is, $\boldsymbol{u} = u \boldsymbol{\hat z}$, (\ref{static:1}) and (\ref{dynamic:1}) can be rewritten as $m g = \kappa I$ and $\cU = \kappa u$,respectively, where $\kappa=BL$ is a geometric factor. By eliminating it, we obtain the measurement equation $m g u = \cU I$, which, virtually, balances mechanical and electrical powers. To derive this equation, current, magnetic field, and velocity must be assumed as rigorously uniform. These assumptions are also required to describe macroscopically the watt-balance operation by using the gradient of the magnetic flux linked to the coil. In fact, it must be considered that current and field inhomogeneities make, in principle, the fluxes linked in the static and dynamic coil-operation different.

\section{Static phase}

\subsection{Magnetohydrodynamics equations}
In order to estimate the Hall field and charges distribution in the watt-balance coil, we model a metal as proposed  by Goedecke and Kanim \cite{Goedecke:2006}. Hence, we consider two interpenetrating compressible and isothermal charged-fluids -- i) the ions lattice and ii) the plasma of free conduction electrons -- confined in a stationary wire, where forces of constraint are applied to hold the ion lattice in place. As a result, there will be a lattice deformation and a change of the electron density. In a steady state, the fluid equations,
\numparts \begin{eqnarray}
 \bnabla (n_\rme \bv_\rme) &= 0 \label{continuity} \\
 -n_\rme e (\bE + \bv_\rme \times \bB) - \bnabla p_\rme - \rho (n_\rme e)^2 \bv_\rme &= -\mu_\rme \bi{g}
 \label{electrons} \\
  n_\rmi e \bE - \bnabla p_\rmi + \rho (n_\rme e)^2 \bv_\rme &= -\mu_\rmi \bi{g} \label{ions}
\end{eqnarray} \endnumparts
are the continuity equation (\ref{continuity}), the momentum transfer equations (\ref{electrons}) and (\ref{ions}), and the equations of state
\numparts \begin{eqnarray}
 \bnabla p_\rmi &= (\partial_{n_\rmi} p_\rmi)_0 \bnabla n_\rmi &= \frac{\cK_\rmi \bnabla n_\rmi}{n_\rmi} \\
 \bnabla p_\rme &= (\partial_{n_\rme} p_\rme)_0 \bnabla n_\rme &= \frac{\cK_\rme \bnabla n_\rme}{n_\rme} .
\end{eqnarray} \endnumparts
Since the free-electron and ion number-densities,
\numparts \begin{eqnarray}
 n_\rme &= n_0(1+\zeta_\rme) \label{ze} \\
 n_\rmi &= n_0(1+\zeta_\rmi) \label{zi} ,
\end{eqnarray} \endnumparts
deviate only by the small amounts $\zeta_{\rme,\rmi}$ from the mean value $n_0$, $(\partial_n p)_0$ means $(\partial_n p)_{T,n_0}$. In the equations (\ref{electrons}) and (\ref{ions}), the forces acting on the fluids are: the Lorentz force (the term proportional to $\bB$ is missing for the ions because they are immobile), the pressure gradients, the force due to the electron scattering on the lattice ions, and the gravity. For the sake of simplicity, we assume that the charge carriers are electrons and that there is one charge carrier per ion. In (\ref{continuity}-$c$), $e$ is the elementary charge, $\bv_\rme$ is the electron drift-velocity, the ion drift-velocity is zero, $p_{\rme,\rmi}$ are the electron and ion pressures, $\cK_\rmi$ and $\cK_\rme=\frac{2}{3}\epsilon_F n_0$ are the bulk moduli of the ions and free-electrons, and $\epsilon_F$ is the Fermi energy. According to the Drude model of electrical conduction, the frictional-force density is $\rho (n_0 e)^2 \bv_\rme$, where $\rho$ is the electrical resistivity. The density of the electrical current is $\bj = -n_\rme e \bv_\rme$. We also considered the self weight of the free-electrons and ions; $\bi{g}=-g\hat{\bi{z}}$ is the acceleration due to gravity and $\mu_\rme$ and $\mu_\rmi$ are the free-electron and ion mass-densities. By using (\ref{ze}-$b$) in (\ref{continuity}-$c$), we obtain
\numparts \begin{eqnarray}
 \bnabla \bv_\rme + \bnabla (\zeta_\rme \bv_\rme) &= 0 \label{master}\\
 -n_0 e (\bE + \bv_\rme \times \bB) - \cK_\rme \bnabla \zeta_\rme - \rho (n_0 e)^2 \bv_\rme &= -\mu_\rme \bi{g} \\
  n_0 e \bE - \cK_\rmi \bnabla \zeta_\rmi + \rho (n_0 e)^2 \bv_\rme &= -\mu_\rmi \bi{g}
\end{eqnarray} \endnumparts
where the terms multiplied by $\zeta_\rme$ and $\zeta_\rmi$ have been neglected, leaving only that terms multiplied  by their gradients.

The electric field $\bE=\bE_0+\bcE$ and magnetic flux density $\bB=\bB_0+\bcB$ include the external fields, $\bE_0$ and $\bB_0$, as well as the fields generated by the charge and current distributions, $\bcE$ and $\bcB$. They are stated by the Maxwell equations
\numparts \begin{eqnarray}
 \bnabla \bcE &= n_0 e (\zeta_\rmi - \zeta_\rme) / \epsilon_0 \label{maxwell:1}\\
 \bnabla \times \bcE &= 0 \\
 \bnabla \times \bcB &= -\mu_0 n_0 e\, \bv_\rme \label{maxwell} \\
 \bnabla \bcB &= 0
\end{eqnarray} \endnumparts

\subsection{Equation solution}
\begin{figure}
\centering
\includegraphics[width=60mm]{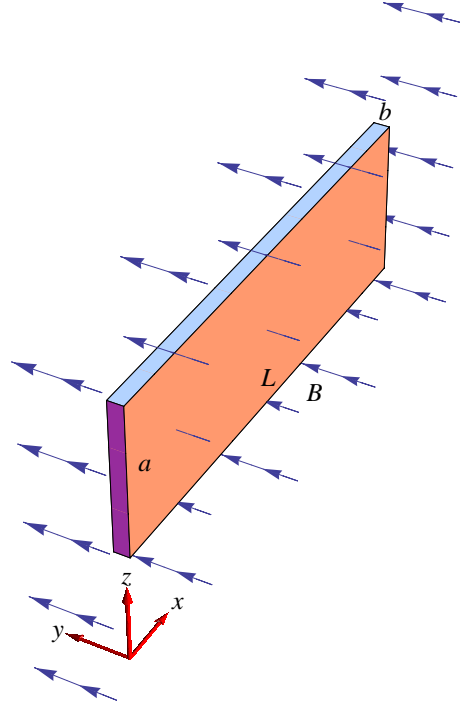}
\centering
\caption{Schematic view of the rectilinear wire in a magnetic field. The wire dimensions are $L$, $b$, and $a$, respectively; $B$ is the magnetic flux density.}\label{scheme}
\end{figure}

As shown in Fig. \ref{scheme}, we consider a rectilinear wire having rectangular $b \times a$ cross-section in the $y-z$ plane and extending from $-L/2$ to $L/2$ in the $x$ direction. An electric current flows in the positive $x$ direction, having a density $\bj=j_0[1+\iota(z), 0, 0]^T$ and an associated drift-velocity of the free electrons $\bv_\rme=-\bj/(n_\rme e)$ and external field $\bE_0=[E_0, 0, 0]^T$. The wire is in an external magnetic flux density $\bB_0=[0, B_0, 0]^T$ pointing in the $y$ direction. We assume that all quantities in (\ref{master}-c) and (\ref{maxwell}-d) depend only on $z$; strictly speaking, this corresponds to assume an infinite extension of the wire in the $x$ and $y$ directions, in order to call on invariance arguments.

The magnetohydrodynamics and field equations are
\numparts \begin{eqnarray}\label{sc2}
  E_0 = \rho j_0 \\
  j_0 B_0 - n_0 e \cE_z - \cK_\rme \partial_z \zeta_\rme & = \mu_\rme g \label{WB2} \\
  n_0 e \cE_z - \cK_\rmi \partial_z \zeta_\rmi & = \mu_\rmi g \label{WB3} \\
  \epsilon_0 \partial_z \cE_z - n_0 e (\zeta_\rmi - \zeta_\rme) &= 0
\end{eqnarray} \endnumparts
where the continuity equation (\ref{master}) is identically satisfied and the equation (\ref{maxwell}) has been omitted, i.e. the magnetic field generated by $\bj$ has been neglected. The first equation, expressing the equilibrium of the $x$ components of the forces acting on the free electrons, is the Ohm law. In the case of a cylindrical wire and magnet geometry, the magnetohydrodynamics equations can be solved with equivalent results by using the same approximations.

\begin{figure}
\includegraphics[width=62mm]{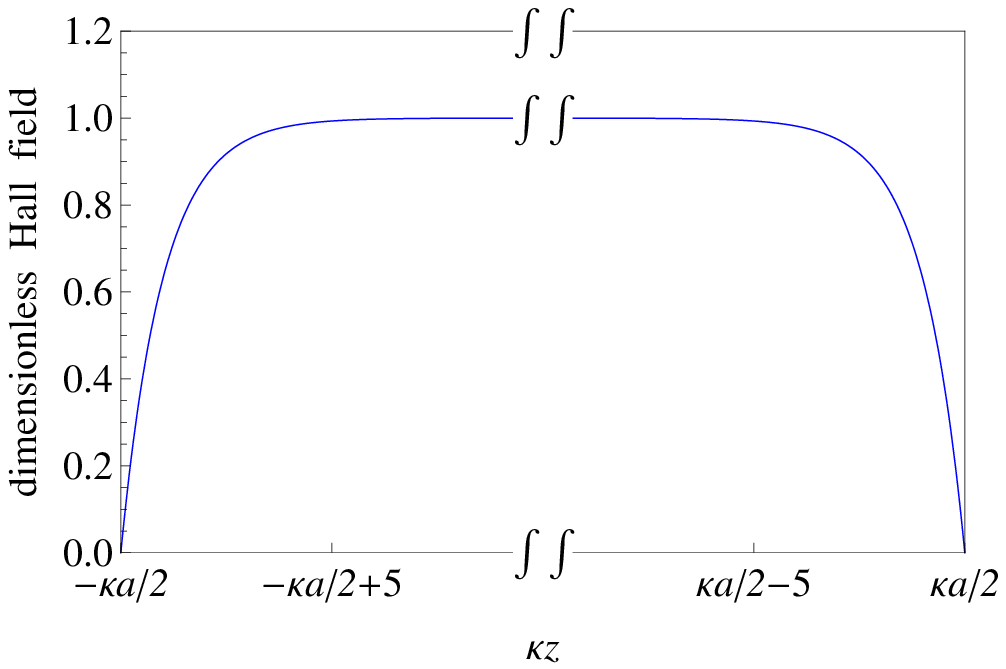}
\includegraphics[width=67mm]{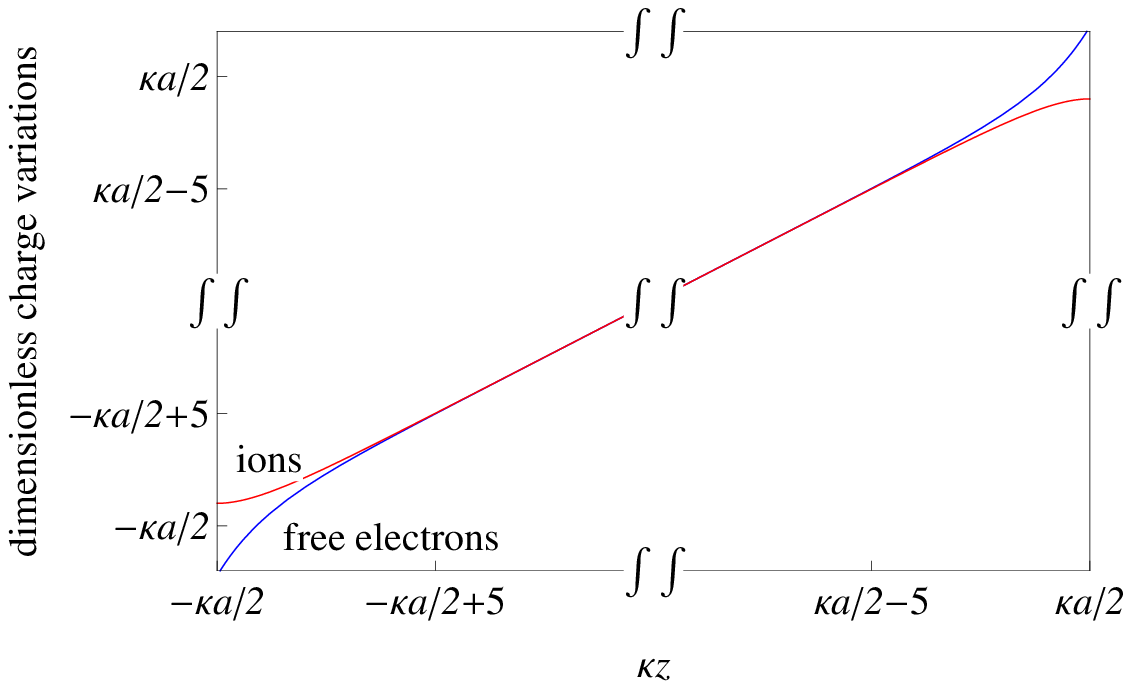}
\caption{Dimensionless Hall field -- $\frac{n_0 e(\cK_\rmi + \cK_\rme)\cE_z}{(B_0 j_0 + {\bar \mu}g) \cK_\rmi}$, left -- and variations of the free-electron and ion densities -- $\frac{\kappa(\cK_\rmi+\cK_\rme)\zeta_{\rme,\rmi}}{B_0 j_0 - \mu g}$, right -- across a wire in the magnetic flux density $B_0$ and carrying the electrical current $j_0$.}\label{Hall}
\end{figure}

To solve (\ref{sc2}-$d$), we impose the boundary conditions
\numparts \begin{equation}\label{boundary-1}
  \int_{-a/2}^{+a/2} \zeta_{\rme,\rmi}(z)\; \rmd z = 0 ,
\end{equation}
which expresses that $n_0$ is the mean number density, and
\begin{equation}\label{boundary-2}
  \cE_z(-a/2) = \cE_z(a/2) = 0,
\end{equation} \endnumparts
because the wire has no net charge and it is assumed to extend to the infinity in the $x$ and $y$ directions. With these boundary conditions, the solutions of (\ref{sc2}-$d$) are
\numparts \begin{eqnarray}\label{hall-solution}
     \cE_z(z) = \frac{(B_0 j_0 + {\bar \mu}g) \cK_\rmi }{n_0 e(\cK_\rmi + \cK_\rme)}
     \bigg[ 1 - \frac{\cosh(\kappa z)}{\cosh(\kappa a/2)} \bigg] , \\ \label{hall-solution-b}
     \zeta_\rme(z) = \frac{B_0 j_0 - \mu g}{\kappa(\cK_\rmi+\cK_\rme)}
     \left[ \kappa z + \frac{\cK_\rmi \Xi_0 \sinh(\kappa z)}{\cK_\rme \cosh(\kappa a/2)} \right], \\
     \zeta_\rmi(z) = \frac{B_0 j_0 - \mu g}{\kappa(\cK_\rmi+\cK_\rme)}
     \left[ \kappa z - \frac{\Xi_0 \sinh(\kappa z)}{\cosh(\kappa a/2)} \right] ,
\end{eqnarray} \endnumparts
where
\numparts \begin{equation}
 \kappa = \sqrt{ \frac{\cK_\rme + \cK_\rmi}{\cK_\rme \cK_\rmi} } \frac{n_0 e}{\sqrt{\epsilon_0}} ,
\end{equation}
\begin{equation}
 \Xi_0 = \frac{ B_0 j_0 + {\bar \mu}g }{B_0 j_0 - \mu g} ,
\end{equation}
\begin{equation}
 \bar \mu = \frac{\mu_\rmi \cK_\rme - \mu_\rme \cK_\rmi}{\cK_\rmi} ,
\end{equation}\endnumparts
and $\mu=\mu_\rmi+\mu_\rme$. The dimensionless -- in square brackets -- Hall field and the free-electron and ion densities across the wire are shown in Fig.\ \ref{Hall}; the numerical values of the model parameter are given in table \ref{num-values}.

Equation (\ref{hall-solution}) shows two modifications of the textbook equation $\cE_z=B_0j_0/(n_0 e)=R_H B_0 j_0$, where $R_H$ is the Hall coefficient \cite{Ashcroft:1976}. The first and most important modification, scales $R_H$ by the ratio of the bulk modulus of the ions to the sum of the bulk moduli of the free-electron and ions, $\cK_\rmi / (\cK_\rmi + \cK_\rme)$. Furthermore, $\cE_z$ depends on the depth in the wire (see Fig.\ \ref{Hall}, left). Neglecting this $z$ dependence, which is relevant only within sheaths whose thickness is of the order of $\kappa^{-1}\simeq10^{-10}$ m, the Hall coefficient is $R_H=1/(n_0 e)$ only if $\cK_\rme/\cK_\rmi \rightarrow 0$. The opposite $\cK_\rmi/\cK_\rme \rightarrow 0$ limit yields a zero Hall coefficient. Since the ion bulk modulus is not so larger than the electron-gas one, the modification predicted by (\ref{hall-solution}) is significant. The reason why $\cE_z$ depends on the $\cK_\rme/\cK_\rmi$ ratio is that the Lorentz force is counteracted by both $\cE_z$ and the electron-gas stiffness; a low stiffness gives this task to the Hall field, a high stiffness does not require the Hall field contribution. As regards the ion lattice, by examining (\ref{WB3}), we observe that a null stiffness imposes a null Hall field whereas an infinite stiffness leads to the maximum Hall field.

According to the textbook formula, for monovalent transition metals as Cu, Ag and Au, the product $R_H n_0 e$ should be identically one. However, the experimentally determined values are smaller than the unity, as shown in table \ref{tab:RH}, the corrective term $\cK_\rmi / (\cK_\rmi + \cK_\rme)$ accounts for this discrepancy, though less effectively for gold.

\begin{table}[b]
\caption{\label{tab:RH}Measured Hall coefficients $R_H n_0 e$ compared to the ones estimated by (\ref{hall-solution}). The measured values are from \cite{Ashcroft:1976}.}
{\small
\begin{tabular}{ccccccc}
\br
\textbf{Element} &$\epsilon_F$/eV &$10^{28}n_0/$m$^{-3}$ &$K_i$/GPa &$K_e$/GPa &$R_H n_0 e$ &$\cK_\rmi / (\cK_\rmi + \cK_\rme)$ \\
\br
\textbf{Cu} & 7.10 & 8.49 & 140 & 63.4 & 0.67 & 0.68 \\
\textbf{Ag} & 5.50 & 5.86 & 100 & 34.4 & 0.77 & 0.74 \\
\textbf{Au} & 5.53 & 5.90 & 180 & 34.8 & 0.67 & 0.84 \\
\br
\end{tabular}
}
\end{table}

The second, minor, modification accounts for the different effects of self-weigh on the electron gas and ion lattice; also in the absence of the magnetic field, an electric field originates across the conductor. In fact, since the electron-gas and ion stiffnesses are different, the differential strain due to gravity split the positive and negative charge distributions and generates an electric field. Owing to the layout of a watt balance, this field and the Hall fields are collinear.

As sown in Fig.\ \ref{Hall} (right), both the free-electron and ion densities increases linearly along the $z$ direction. These increases correspond to the strains caused by the compressive stresses originated by the Lorentz force (free electrons) and the Hall field (ion lattice); opposite strains are induced by the gravity. At the $z=\pm a/2$ surfaces, charge layers are created that are the sources of the Hall field.

\subsection{Magnetic force}
In (\ref{sc2}-$d$) the mean current-density $j_0$ is an external constraint. The simplifications made do not allow the current distribution $\iota(z)$ to be determined. However, by twisting a bit the mathematical rigor, at least for an order-of-magnitude estimate, we can use the Drude model of electrical conduction and assume that the $x=$ const. end faces of the wire are equipotential to write the current density as $j_0(1+\zeta_\rme)$. Therefore, the magnetic force is
\begin{equation}\label{Force}
 F = B_0 L I_0 \left[ 1 + \frac{1}{a} \int_{-a/2}^{+a/2} \beta(z) \zeta_\rme (z) \,\rmd z \right] ,
\end{equation}
where $B_0(1+\beta)$ is the magnetic flux density, $B_0$ is the mean flux density, and $I_0=ab j_0$ the current flowing in the wire. Since $\zeta_\rme (z)$ is an odd function, only the vertical gradient $B_{,z}=\partial_z B$  contributes to (\ref{Force}); hence,
\begin{equation}\fl\label{correction}
 F \approx B_0 L I_0 \left[ 1 + \frac{B_{,z}}{a} \int_{-a/2}^{+a/2} z \zeta_\rme (z) \,\rmd z \right] \approx
 B_0 L I_0 \left[ 1 + \frac{a^2(B_0 j_0 - \mu g)B_{,z}}{12(\cK_\rme+\cK_\rmi) } \right] .
\end{equation}

This equation predicts a correction to the $B_0 L I_0$ expression of the magnetic force. In the case of a typical watt-balance experiment, the magnetic force is of the order of 10 N. Hence, with a single copper ring of $L=1$ m circle and a flux density $B_0=1$ T, the required current is 10 A; if the dimensions of the wire cross-section are $a=10$ mm and $b=1$ mm, the current density is $j_0=1$ A/mm$^2$. A pessimistic estimate of the vertical gradient of the magnetic flux density is $B_{,z}=1$ T/m. With the numerical values of the remaining model parameters summarized in Table \ref{num-values}, the associated relative correction is $4 \times 10^{-11}$ -- given or taken a ten percent due to the gravitational load $\mu g$. Thankfully, this figure makes the correction irrelevant to the watt-balance operation.

\begin{table}
\caption{\label{num-values}Numerical values of the model parameters in the case of a copper wire.}
{\small
\begin{tabular}{llll}
\br
$\epsilon_0 = 8.9 \times 10^{-12}$ F/m &$\epsilon_F = 7$ eV &$g = 9.81$ m/s$^2$ &$\rho = 1.56 \times 10^{-8}$ $\Omega$ m \\
$\cK_\rmi = 140$ GPa &$\cK_\rme = 63$ GPa &$n_0 = 8.5\times 10^{28}$ m$^{-3}$ &$\kappa = 2.2 \times 10^{10}$ m$^{-1}$ \\
$\mu = 8960$ kg/m$^3$ &$\bar{\mu} = 4060$ kg/m$^3$ &$B_0 = 1$ T            &$\partial_z B = 1$ T/m \\
$L = 1$ m &$a = 10^{-2}$ m &$j_0 = 10^6$ A/m$^2$ \\
\br
\end{tabular}
}
\end{table}

\subsection{Extension to a coil of wire}
A last detail must be examined. In practice, the balance coil is made by winding up many wire turns and, owing to the ohmic voltage drop, a potential difference is set up between subsequent turns. Therefore, the boundary condition (\ref{boundary-2}) becomes
\begin{equation}
  \cE_z(-a/2) = \cE_z(a/2) = E_d ,
\end{equation}
where $E_d$ is the electric field in the gap between subsequent coil-windings, and the solutions of (\ref{sc2}-$d$) become
\numparts \begin{eqnarray}
     \cE_z(z) = \frac{(B_0 j_0 + {\bar \mu}g) \cK_\rmi }{n_0 e(\cK_\rmi + \cK_\rme)}
     \bigg[ 1 - \frac{\Xi_1 \cosh(\kappa z)}{\cosh(\kappa a/2)} \bigg] ,\\
     \zeta_\rme(z) = \frac{B_0 j_0 - \mu g}{\kappa(\cK_\rmi+\cK_\rme)}
     \left[ \kappa z + \frac{\cK_\rmi \Xi_2 \sinh(\kappa z)}{\cK_\rme \cosh(\kappa a/2)} \right], \\
     \zeta_\rmi(z) = \frac{B_0 j_0 - \mu g}{\kappa(\cK_\rmi+\cK_\rme)}
     \left[ \kappa z - \frac{\Xi_2 \sinh(\kappa z)}{\cosh(\kappa a/2)} \right] ,
\end{eqnarray} \endnumparts
where
\begin{equation}
 \Xi_2 = \frac{ B_0 j_0 + {\bar\mu}g - {\bar q}E_d }{B_0 j_0 - \mu g} .
\end{equation}\endnumparts
\begin{equation}
 \Xi_1 = \frac{B_0 j_0 + {\bar\mu}g - {\bar q}E_d}{B_0 j_0 + {\bar\mu}g} ,
\end{equation}
and
\numparts\begin{equation}
 {\bar q} = \frac{n_0 e(\cK_\rme+\cK_\rmi)}{\cK_\rmi} ,
\end{equation}

As regards the electron density, the only change with respect to (\ref{hall-solution-b}) is the additional term ${\bar q}E_d$ in the new scale factor $\Xi_2$ of the charge sheaths, which is necessary to ensure that the only internal field is the Hall's one. Since the surface charge-layers do not contribute significantly to the integral in (\ref{correction}), the given expression for the magnetic-force correction is unchanged. A correction reduction of many order of magnitudes is consequential to the reduction of the wire height.

\section{Dynamic phase}

\subsection{Solution of the magnetohydrodynamics equations}
\begin{figure}
\includegraphics[width=64mm]{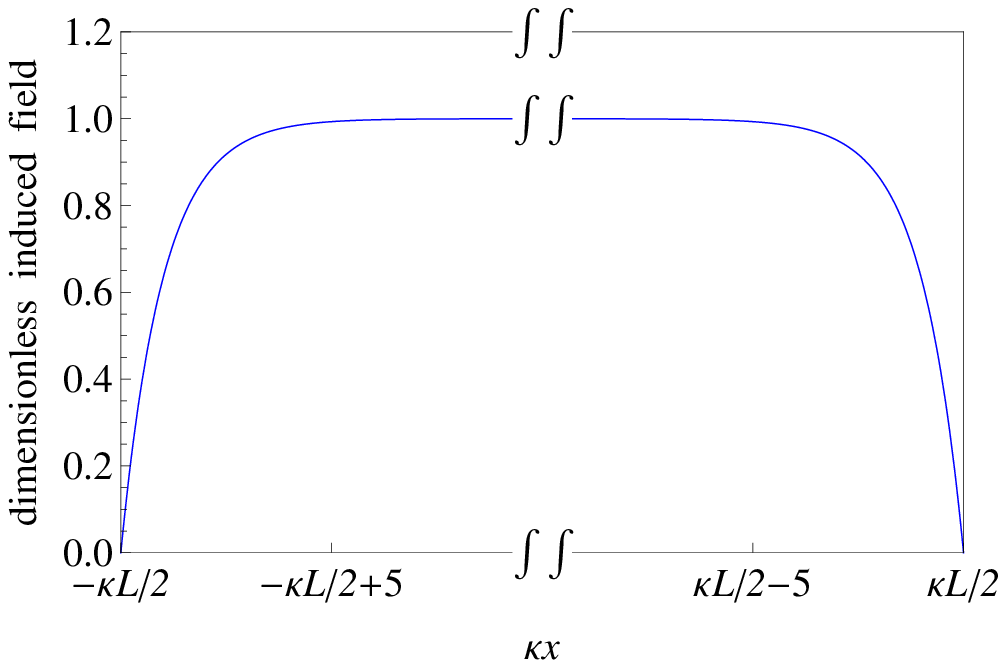}
\includegraphics[width=66mm]{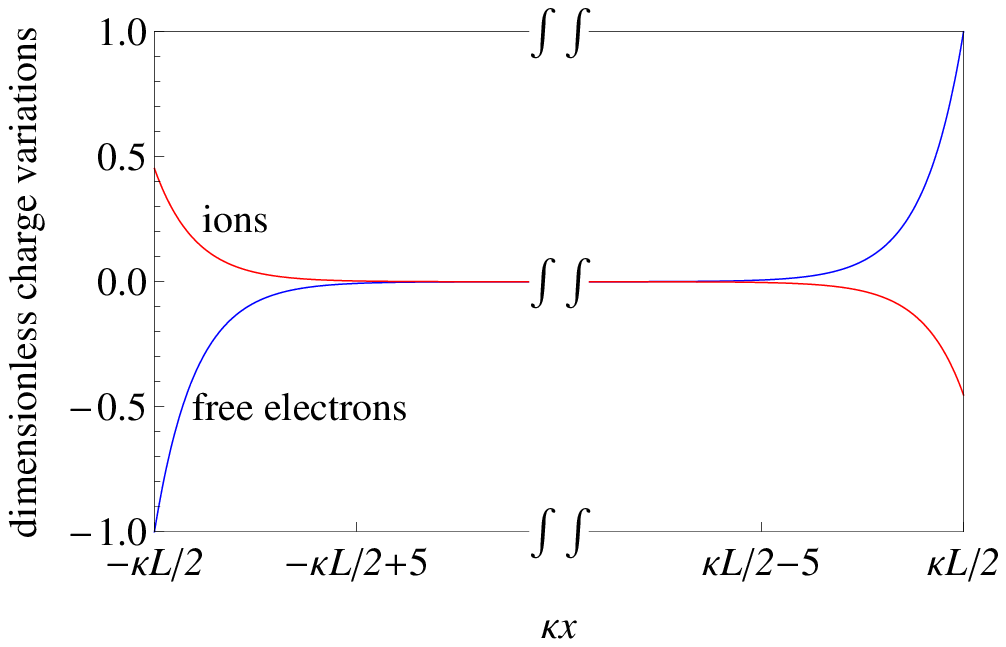}
\caption{Dimensionless electric field -- $\frac{\cE_x}{B_0 u}$, left -- and free-electrons and ions densities -- $\frac{\kappa \cK_{\rme,\rmi} \zeta_{\rme,\rmi} }{n_0 e u B_0}$, right -- along a wire moving in the magnetic flux density $B_0$ with velocity $u$.}\label{fig-static}
\end{figure}

Let us now consider the coil moving in the vertical direction with a constant velocity $u$. The wire is assumed to extend up to the infinity in $x$ direction, but the magnetic field is zero outside the $[-L/2, +L/2]$ interval. In addition, we assume that all the relevant quantities depend only on $x$. The magnetohydrodynamics and field equations are
\numparts \begin{eqnarray}\label{dc2}
  n_0 e(u B_0 - \cE_x) - \cK_\rme \partial_x \zeta_\rme & = 0 \label{WB22} \\
  n_0 e(\cE_x - u B_0) - \cK_\rmi \partial_x \zeta_\rmi & = 0 \label{WB32} \\
  \epsilon_0 \partial_x \cE_x - n_0 e (\zeta_\rmi - \zeta_\rme) &= 0
\end{eqnarray} \endnumparts
where
\numparts \begin{equation}
  \int_{-L/2}^{+L/2} \zeta_{\rme,\rmi}(x)\; \rmd x = 0 ,
\end{equation}
because $n_0$ is the mean number density in the $[-L/2, +L/2]$ interval, and
\begin{equation}
  \cE_x(-L/2) = \cE_x(L/2) = 0,
\end{equation} \endnumparts
because outside the $[-L/2, +L/2]$ interval the electric potential is constant. With these boundary conditions, the solutions of (\ref{dc2}-$d$) are
\numparts \begin{eqnarray}\label{d-solution}
     \cE_x(x) = u B_0 \left[ 1 - \frac{\cosh(\kappa x)}{\cosh(\kappa L/2)} \right] , \\
     \zeta_\rme(x) =  \frac{n_0 e u B_0}{\kappa \cK_\rme}\frac{\sinh(\kappa x)}{\cosh(\kappa L/2)} , \\
     \zeta_\rmi(x) = -\frac{\cK_\rme}{\cK_\rmi} \, \zeta_\rme(x) .
\end{eqnarray} \endnumparts
The dimensionless -- in square brackets -- induced field and the free-electron and ion densities along the wire are shown in Fig.\ \ref{fig-static}; the numerical values of the model parameters are given in table \ref{num-values}. As shown in Fig.\ \ref{fig-static} (right), the free-electron and ion lattice are shifted in opposite directions to generate charge layers at the $\pm L/2$ ends of the wire, which are the sources of the induced electric field.

\subsection{Induced electric potential}
The induced electric potential is obtained by integrating $\cE_x(x)$ along the wire axis. Actually, this integration can be confined within the interval $[-L/2,L/2]$, because outside this interval there is no magnetic field to generate a Lorentz force counteracting an electric field. Hence,
\begin{equation}\label{emf}
 \cU = \int_{-L/2}^{+L/2} \cE_x(x) \, \rmd x = u B_0 L \left[ 1 - \frac{2 \tanh(L\kappa/2)}{\kappa L} \right] .
\end{equation}
As (\ref{correction}) does for the magnetic force, (\ref{emf}) predicts a correction to the textbook expression of the induced electric potential. The predicted potential is smaller than $uB_0 L$; as shown in Fig.\ \ref{fig-static} (left), the reduction originates in the smooth transition of $\cE_x(x)$ from the bulk value $uB_0$, in the conductor part immersed in the magnetic field, to zero, in the part external to the field. With the parameter values given in table \ref{num-values} and a single winding of $L=1$ m length, the relative correction is $9.2 \times 10^{-11}$. Also in this case, the correction is irrelevant to the watt-balance operation. In a real coil, made by winding up many wire turns, the much greater $L$ value ensures that this correction is further scaled down by orders of magnitude.

\section{Conclusions}
The basic principles of operation of the watt balance experiment have been examined from a microscopic viewpoint in order to verify the presence of corrective terms to the formulae for the magnetic force acting on a conductor in a magnetic field and for the electric potential induced by its motion. The model used describes the conductor as a plasma with interpenetrating compressible charged fluids: an ion lattice and a free-electron gas. In order to solve analytically the relevant magnetohydrodynamics equations an extreme simplification of the coils-magnet system is necessary, but the order-of-magnitude of the sought corrections should have been correctly estimated. These corrections are many order-of-magnitudes below the sensitivity of any present and future watt-balance experiment. Nevertheless, they shed light in the background of the watt balance operation.

\ack
This work was jointly funded by the European Metrology Research Programme (EMRP) participating countries within the European Association of National Metrology Institutes (EURAMET) and the European Union.

\end{document}